\documentstyle[preprint,aps]{revtex} 
\begin{document}
\draft
\title{\bf{Nature of eigenstates in a mesoscopic ring coupled to
a side branch}} 
\author{P. A. Sreeram \cite{eml2} and P. Singha Deo \cite{eml1} }
\address{Institute of Physics, Bhubaneswar 751005, India}
\maketitle
\begin{abstract}
We formalize the parity effect of a multichannel ring coupled to a
side branch using the tight binding model. We find that the tight
binding model gives slightly different result from the continuum model.
We also show that some states of the system can be nonmagnetic.
In the multichannel ring coupled to a
side branch, 
the persistent current in the ring
does not change sign,
over gigantic fluctuations in N (the 
number of particles in the ring). 
\end{abstract}
\narrowtext
\newpage

{\bf I Introduction}

Prior to the experiments \cite{1,2,3},
Buttiker et al \cite{4} had shown the possibility of
observing equilibrium persistent currents in normal metal or semiconductor
rings pierced by a magnetic flux $\phi$.
The eigenstates of a quantum ring are flux dependent and as the flux is
varied, one gets a E versus $\phi$ dispersion curve. States whose E versus 
$\phi$ dispersion curve have positive slope carry diamagnetic current
and states whose E versus $\phi$ dispersion curve have negative slope carry
paramagnetic current. 
For free electrons in a single 
channel ring consecutive eigenenergies have opposite slope and the magnetic
response of the ring is diamagnetic for an odd number of electrons and 
paramagnetic for an even number of electrons in the ring \cite{5}. This is the 
essence of the parity effect. It was shown by Leggett that this effect is true
for any one body and two body scattering and arises just because of the
antisymmetric nature of the many body fermionic wavefunction \cite{6}. 
With the addition of an extra electron in the ring, there is a statistical
phase that shifts the E versus $\phi$ dispersion curve by $\phi_0/2$ and hence
the N body and N+1 body states carry opposite persistent currents \cite{7}.
The parity effect is not destroyed by
spin, finite temperature or disorder \cite{5,7}. 
Interactions do not destroy the parity effect but interactions with
spin can lead to the creation of a fractional Aharonov- Bohm effect, which
has however not been observed experimentally yet 
\cite{8,9,10,11,12,13,14,15,16}.
The parity effect is
also observed in multichannel simulations \cite{17}. 
However, there the essence of
parity effect is slightly modified. 
When there is complete rotational symmetry (clean multichannel ring), subband
quantum number is a good quantum number and consecutive states belonging to 
the same subband have opposite slope. If the rotational symmetry is destroyed
(say, by impurities),
then different subbands can mix. But this only opens up a gap at points where
states cross (i.e. it lifts degeneracy) and flattens the E versus 
$\phi$ dispersion
curves, but does not change the sign of the slope. 
As a result, what is seen
in a multichannel ring is that the states change slope at the scale of 
Thouless energy.
 So the parity dominated current 
of a single ring has an {\em a priori} random sign depending on the number of 
electrons present in it and this allows us to treat a collection of rings 
\cite{1}
as a statistical ensemble. Various models have been used
to calculate the ensemble average \cite{17,18,19,20}, which is a nontrivial
task. In one such model, one assumes that the number of electrons N in all 
the rings is the same (strong canonical ensemble) and averaging is done over
impurity configurations.

The system of a ring coupled to a side branch has recently attracted 
attention \cite{21,22,23,24}. So far it is studied using a 1D modeling. 
In this model
the geometry consists of a 
finite wire or a side chain,
of length $v$ being attached to a ring of length $u$, as
shown in Fig (1). Ref. \cite{21} points out 
some interesting aspects of the electronic states in a quantum ring
weakly coupled to a side branch.
Potential scattering at the junction of the ring 
and the side branch can cause such weak coupling of the states. 
On the other hand, ref \cite{22} studies
the states of the combined system of ring coupled to a side branch,
from first principles without any additional
potential scattering at the junction. 
In this strong coupling regime, one
cannot distinguish the states of the ring and the side branch separately
because they carry the same amount of persistent current. However,
one can clearly observe the effect of the two length scales ( $v$ and $u$) on 
the eigenstates of the system. It was shown that the states associated 
with the energy scale $v$ do not show a parity effect. If there are $n$ states
associated with the length scale $v$ between two states associated with the
length scale $u$ then all the $n$ states have the same slope. 
Ref \cite{24}
studies 
the effect of charge transfer by polarizing
the side branch with respect to the ring. Under such charge transfer, one can
measure the capacitance coefficients of the ring \cite{21,24}.
The flux dependence of the eigenenergies determine the flux dependence of
the capacitance coefficients.
As magnetization measurements \cite{1,2,3} are a difficult task, 
one can alternately
measure the flux dependent capacitance coefficients of the ring, 
to
probe it's equilibrium properties.
Ref \cite{25} presents an interesting study of
coupled quantum rings. 

For the first time, we try to analyze 
the flux dependence of the eigenenergies of a multichannel
ring coupled to a multichannel side branch. We restrict ourselves to the
free electron case, but, 
following the argument of Leggett, electron-electron interaction will not 
change this 
parity effect observed for free electrons \cite{6,23} and 
this has been rigorously
established in the framework of a Luttinger Liquid \cite{9,10}.
Also it was proposed in 1D \cite{22}, that by fabricating such a geometry, 
with a large $v\over{u}$
(say, 10) one can get rid of the parity effect and hence the tough problem
of ensemble averaging in a many ring experiment. In fact, the strong
canonical ensemble may be a good approximation to estimate the disorder
averaged persistent current, if for small fluctuations in N, the persistent 
current does not change sign.
The effect that was shown in 1D should survive
multichannel effects if it is to be of experimental relevance. Besides,
the conductance across such a system  (side branch coupled to a ring) 
was measured in a recent experiment
\cite{26} in the Coulomb blockade regime 
and the conductance features are believed to depend on the parity of the 
eigenstates of the closed system \cite{27}.

{\bf II Theoretical Treatment}

To study
the multichannel situation, it is sufficient to consider a 2D geometry.
One of the easiest ways of treating a multichannel ring is to have a ring
made up of sites, the system being described by the tight-binding Hamiltonian.
The system thus considered is shown in Fig (2).
If $a$ is the lattice parameter, then the length of the ring is $L_{r}a$, width
of the ring is $w_r a$, length of the side chain is $L_s a$ and width of the
side chain is $w_s a$.
The Hamiltonian describing the system can be written as,
\begin{equation}
H=-\sum_{jk}\left [ \delta t_{jk}^\alpha a_{jk}^\dagger a_{j+1 k} + t_{jk}^\beta 
a_{jk}^\dagger a_{j k+1} + h.c. \right ]
\end{equation}
Here, $\delta$ = $e^{i 2\pi \phi\over L_r \phi_0}$, where $\phi$ is the flux
threading the 2D 
ring, $\phi_0$ = ${hc\over e}$, the elementary flux quantum. 
a$^\dagger_{jk}$ is
the electron creation operator at the site (j,k). 
The
first term in H represents hopping in the azimuthal direction which is the propagating 
direction ($\alpha$ being the index for it). The second term represents
hopping in the transverse direction. We have considered $t^{\alpha}$ =
$t^{\beta}$ = t and only
nearest neighbour 
hopping.
We perform an exact diagonalization of the Hamiltonian to evaluate the 
single particle levels.
The persistent current carried by a particular 
level $E_{n}$ is given by, (we have set $\hbar$ = 2 m = c = 1)

\begin{equation}
 I_n=-{ \partial  E_n \over {\partial \phi}}
\end{equation}
 \noindent and the total persistent current is just the sum 
of the persistent currents
 of all the filled levels. In presence of spin each state will be filled
by two electrons but the effect of spin is not the subject of the present 
study.
The probability that a level is filled is given
 by the Fermi factor. We restrict ourselves to finding the persistent current
 at zero temperature. 

Using the continuum model in a multichannel ring is very 
hazardous. So far we know that the two models (continuum and discrete)
give results that agree
 with each other. But we find that if the ring has a finite side chain
 attached to it, as is the case considered by us then the two models
 give qualitatively the same result as in 1D but there are some major
quantitative differences.
A free electron in an infinite 1D line behaves like an electron in the 
band of an infinite  1D
periodic system. But the finite size effects in the two cases are
always different, i.e. a free electron in a finite 1D line with some 
boundary conditions at the two ends of the line is expected to behave 
differently than an electron in a finite lattice. A ring has no boundaries
and 
is effectively an infinite system. However, if we attach a finite side chain
to the ring, then putting sites in it or not putting sites in it will
always make a difference. As the electron is moving in a background of
positively charged ions, it may feel a periodic potential and then the tight
binding model may be the more appropriate description. On the other hand, if
the number of sites is very large then it approaches the
continuum 
limit. First we will establish that the two
models give qualitatively the same result in the 1D regime. This will justify
studying the
multichannel regime using the tight binding model. 

{ \bf III Results and Discussions}

To formalize the study of the parity effect for a tight binding ring 
we first consider
a 1 D ring described by tight binding Hamiltonian.
In Fig (3) we show the typical spectrum of 
E versus $\phi$ dispersion curve of a ring made up of 5 sites to which
a side chain of 10 sites is attached. 
The slope of states change after every three states.
The continuum analogue of this system is a stub of length $v$ attached to
a ring of length $u$ such that $v\over{u}$ = 2. We calculate the E versus $\phi$
dispersion curve of this system using the method of ref \cite{22} 
and the spectrum
is shown in Fig (4). In this case the slopes change after every two states
in contrast to three in the tight binding model. We have checked for various
other cases and for large values of $v\over{u}$, and we find that the 
number of consecutive states with the same sign of slope in the tight
binding model and continuum model at most differ by one.
If there are 10 sites in the stub, then the stub contributes 10 states to the
system and along with the 5 sites of the main wire we have 15 states in all.
If the stub were absent, then, the system would have 5 states with
consecutive states having opposite slope and so the slope would have
changed 5 times in all. When the stub is attached, the 15 states are found
to change their slope exactly 5 times. This makes the slope change after 3
consecutive states. Hence, the additional states created by the side
chain in the tight binding model are like states associated with the length
scale $v$ and they all
are parity violating states. So, the continuum model and the tight binding
model give qualitatively identical results.
Only,
the number of states created by the tight binding
side chain may not be the same as that
of the continuum case. 

When the ring is detached from the side branch, we can assign a definite quantum
number `$s$' to the states of the side branch. And similarly, the states of the ring
can be assigned  a definite quantum number `$r$'. The parity effect of the 
states of the ring can be understood according to Leggett's conjecture
from the antisymmetric property of the many body wavefunction. When
the ring and the side branch are coupled, the states that leak into the ring from
the side branch become magnetic. But states with quantum numbers `$s$' and
`$s+1$' do not have
opposite slopes. However, this too can be understood from the antisymmetric
property of the many body wavefunction. 
The statistical phase is neutralized
by another phase $\pi$ that originates from geometric scattering by 
the side branch \cite{23}.
The difference in phase acquired by an electron in two such consecutive 
states in going round the ring once does not depend on the kinetic energy 
difference of the 
electron in the two states alone but also on the special phase $\pi$.
Note that in ref \cite{16}, Haldane has shown that interactions can give
an additional phase $\pi$ that can neutralize the statistical phase.

An interesting quantum effect that has no classical analog sometimes arises 
if the 
total number of states in the combined system of the ring and the side branch, 
is not an integer times the number
of states of the isolated ring. For example if
the ring is made up of three sites and the side branch is made up of 
two sites then
there are five states in all. 
The eigenvalues of this system can be found analytically by diagonalizing
the Hamiltonian and the 5 eigenvalues $e_1, e_2, e_3, e_4, e_5$ in the range
$0 < \alpha (=\frac{2 \pi \phi}{\phi_0 L_r}) < 2 \pi$ are given
below :

$e_1=\frac{c_1 exp(i \alpha)}{x(\alpha)}+\frac{exp(-i \alpha) x(\alpha)}{18^{1/3}}$

$e_2=-1$

$e_3=i \frac{exp(-i \alpha) y(\alpha)}{2^{5/3} 3^{5/6} x(\alpha)}$

$e_4=1$

$e_5=-e_3$   

where, $x(\alpha)=(-9-9 exp(6i\alpha)+\sqrt{3}(27-202 exp(6i\alpha)+27
exp(12i\alpha))^{1/2})^{1/3}$ ,

$y(\alpha)=-24exp(2i\alpha)+8i\sqrt{3}exp(2i\alpha)+
i2^{1/3}3^{1/6}x^2(\alpha) + 18^{1/3}x^2(\alpha)$,

$c_1=(\frac{128}{3})^{1/3}$.

\noindent Note that $e_2$ and $e_4$
do not depend on $\phi$.
Physically these two states remain completely localized inside the
side branch although there is no potential restricting it from leaking out.
If it leaks out then the breakdown of time reversal symmetry necessitates it
to be magnetic (ie. carry a persistent current) whereas the antisymmetric
property of the many body wavefunction does not allow the five states to change
their slope more than three times (the ring being made up of three sites),
maintaining the symmetry of the tight 
binding band. So, an interplay of these symmetry
principles, keep this state localized to the stub. 
The state forms a node at the junction of the ring and the side branch and the
wavefunction vanishes inside the ring. These are the very states which
give rise to a zero in the transmission if the ring is severed at a point,
such that the system becomes the same as the T shaped stub  of ref \cite{28}. 
It was shown
that the zero in the transmission was a consequence of unitarity \cite{28}.
We argue from symmetry principles, that such states will exist which vanish
in the ring and therefore will be nonmagnetic.

Now, realistic systems are not 1D and there are other length scales 
apart from $L_r (\equiv u)$ and $L_s (\equiv v)$, 
like w$_r$ and w$_s$ that determine the number of states
associated with the ring and the stub. 
Having justified the study of
the parity effect using the tight binding model we proceed to explore 
the parity of the states associated with the length scale $w_s$ of a 
realistic multichannel ring, using the tight binding model. 
We consider the 2D geometry
as shown in Fig (2). 
It is difficult to infer
about the parity effect of a multichannel ring by looking at E versus $\phi$ 
dispersion curves as in Fig (3). Things are more transparent if we look at
the I(N) versus N curve, 
where I(N) is the total persistent current in the ring in units of ($\Gamma$=
$\frac{2 \pi c}{t \phi_0}$)
when there are N electrons in the ring.
In Fig (5), we have chosen $L_r$ = 10, $w_r$ = 5, $L_s$=100 and $w_s$=2 and
plotted $I(N)/\Gamma$ versus N.
The
persistent current does not change sign over long ranges of N.
Most importantly, the number of oscillations 
or number of peaks in $I(N)/\Gamma$ versus N curve is equal to that
of a 10 $\times$ 5 clean ring ( ie. the stub detached from the system of
Fig (5)). 
This shows that all states associated with the stub (or the states associated
with the length scales $L_s$ and $w_s$) 
between two states
of the multichannel ring have the same sign of slope.
As the width of the stub ($w_s$) increases, the stub can accommodate more
and more states within a certain interval of Fermi energy, say, in the
interval of two states of the 10X5 multichannel ring. 
We proceed to study the parity of these states.
Keeping $L_s$, $L_r$, $w_r$ unchanged (ie.
$L_r$=10,$w_r$=5 and $L_s$=100) we increase w$_s$ to 4. The $I(N)/\Gamma$ 
versus N plot
is shown in Fig (6). Then we increase w$_s$ to 6 and the corresponding 
$I(N)/\Gamma$ 
versus N curve is
shown in Fig (7). We find that as w$_s$ increases, over a larger range of N the
persistent current does not change sign. 
Each curve from Fig (5) to Fig (7) looks
similar but magnified.  
The fact that the number of oscillations in the $I(N)/\Gamma$ versus N curve
does not increase signifies that in the multichannel situation the states 
created by the additional length scale $w_s$ of the stub are 
all parity violating states. All the states arising due to the length
scale $w_s$ have the same sign of slope. To compare to what order of
fluctuations in N,
the persistent current changes sign in a multichannel ring we take a ring
that has the same number of propagating modes and the same number of states 
as the system of Fig (5). This
means that we take a ring whose width consists of 5 sites and whose 
length consists
of 50 sites and plot $I(N)/\Gamma$ versus N in Fig (8). The order of oscillations in
the curve compared to that in Fig (5), is prominent. 
Compared to Fig (8), in Fig 5, 6 and 7, there are very broad regions of N
over which the $I(N)/\Gamma$ versus N curve does not change sign. The broadest
region is the central region.
We take the 
central region where say from N = N$_1$ to N = N$_1$ + $\delta$N$_1$ the 
persistent current does not change sign. Then we plot $\delta$N$_1$ versus 
w$_s$ in Fig (9) and find a linear scaling between them. 
One can see that in the case of Fig (7), if the ring is at half filling, even
if N fluctuates by 29 $\%$ of the total number of 
electrons in the ring, 
the sign of the response
will not be $a$ $priori$ random. One should note that the first region where
the persistent currents do not change sign is always diamagnetic but quite 
narrow. The third and fourth regions are quite broad and one can perform a 
many ring experiment populating the ring upto the third or fourth regions.

If we open the ring at a point then we can define a transmission 
amplitude $T$ across the two open ends. 
In the 1D regime, it is the transmission amplitude that relates the outgoing
state to the incoming ones. In the multichannel regime, the outgoing
channels are related to the incoming ones by $T_{mn}$ where 
$T_{mn}$ is the transmission amplitude between the $m$ th subband on one
side and the $n$th subband on the other side. In the
multichannel situation, the discontinuities in $T_{mn}$ \cite{29} will
destroy the parity effect just as  those in $T$ do in 
1D \cite{23}. These discontinuities will give the necessary phase
$\pi$ in the electron wavefunction to neutralize the statistical
phase. This gives rise
to many more states with the same slope. The fact that in the multichannel
situation, every state coming from the additional length scale $w_s$ 
of the stub, in between the states of the ring,
is a parity violating state establish this.

{ \bf IV Conclusion}

So, for the first time we have analyzed in detail the nature of eigenstates
in a ring coupled to a side branch in the multichannel situation. A number of
earlier studies are restricted to the 1D situation. All states associated
with the length scale $w_s$ ( the width of the side branch) between two
states of the ring have the same sign of slope. This is the main result 
of this work. As a result,
even 
for 29 $\%$ fluctuations in N, persistent current in the system may not change
sign. This makes the
system a much better candidate to perform many ring experiments to explore the
finer details of persistent currents without encountering the problem of
ensemble averaging. 
We also show that some states in the 
system can be non-magnetic due to purely quantum mechanical effects. 

{ \bf V Acknowledgments}

One of us (P.S.D) thanks Professor A. M. Jayannavar and we
thank Professor A. M. Srivastava for useful discussions.
\vfill
\eject

\vfil
\eject

\centerline {\bf FIGURE CAPTIONS}

Fig. 1. A stub of length $v$ attached to a ring of length $u$.

Fig. 2. Schematic diagram of a 2D multichannel ring made up of
sites.

Fig. 3. Energy versus $\phi$ dispersion curve for a 1D ring made up 
of 5 sites and
a 1D stub of 10 sites attached to it.

Fig. 4. Energy versus $\phi$ dispersion curve of a 1D ring of 
length `$u$' with a 1D stub of
length `$v$' attached to it for $\frac{v}{u}$=2.

Fig. 5. Persistent Current $\frac{I(N)}{\Gamma}$ in a multichannel 
ring with a multichannel stub attached
to it, plotted as a function of the number of electrons (N). Here, 
$\frac{L_r}{a}$=10,
$\frac{w_r}{a}$=5, $\frac{L_s}{a}$=100, $\frac{w_s}{a}$=2.

Fig. 6. Persistent Current $\frac{I(N)}{\Gamma}$ in a multichannel 
ring with a multichannel stub attached
to it, plotted as a function of the number of electrons (N). Here, 
$\frac{L_r}{a}$=10,
$\frac{w_r}{a}$=5, $\frac{L_s}{a}$=100, $\frac{w_s}{a}$=4.

Fig. 7. Persistent Current $\frac{I(N)}{\Gamma}$ in a multichannel 
ring with a multichannel stub attached
to it, plotted as a function of the number of electrons (N). Here, 
$\frac{L_r}{a}$=10,
$\frac{w_r}{a}$=5, $\frac{L_s}{a}$=100, $\frac{w_s}{a}$=6.

Fig 8. Persistent current I(N) in a clean multichannel ring (no stub attached)
plotted as a function of the number of electrons (N) in the ring.
$\frac{L_r}{a}$=50,
$\frac{w_r}{a}$=5.

Fig 9. Plot of range ($\delta N_1$) of the number of electrons over which
persistent current does not change sign around half filling 
as a function of the stub width
$\frac{w_s}{a}$. 

\vfill
\eject
\end{document}